# An Extremely Elongated Cloud over Arsia Mons Volcano on Mars: II. Mesoscale modeling

**Jorge Hernández Bernal**[1,2], **Aymeric Spiga**[3], **Agustín Sánchez Lavega**[1], **Teresa del Río Gaztelurrutia**[1], **François Forget**[3], **Ehouarn Millour**[3]

1. Dpto. Física Aplicada, EIB, Universidad País Vasco UPV/EHU, Bilbao, Spain
2. Aula EspaZio Gela, Escuela de Ingeniería de Bilbao, Universidad del País Vasco UPV/EHU, Bilbao, Spain
3. Laboratoire de Météorologie Dynamique/Institut Pierre Simon Laplace (LMD/IPSL), Sorbonne Université, Centre National de la Recherche Scientifique (CNRS), École Polytechnique, École Normale Supérieure (ENS), Paris, France

**Corresponding author: Jorge Hernández Bernal, jorge.hernandez@ehu.eus**

## Highlights

- We performed mesoscale model dynamic simulations of the Arsia Mons Elongated Cloud observed in the martian southern solstice.
- Topography-induced circulation causes temperatures to drop by about 30K at the observed origin location and local time of the cloud.
- The cloud tail is much more elongated in the observations than in the model, which challenges our understanding of winds and microphysics.

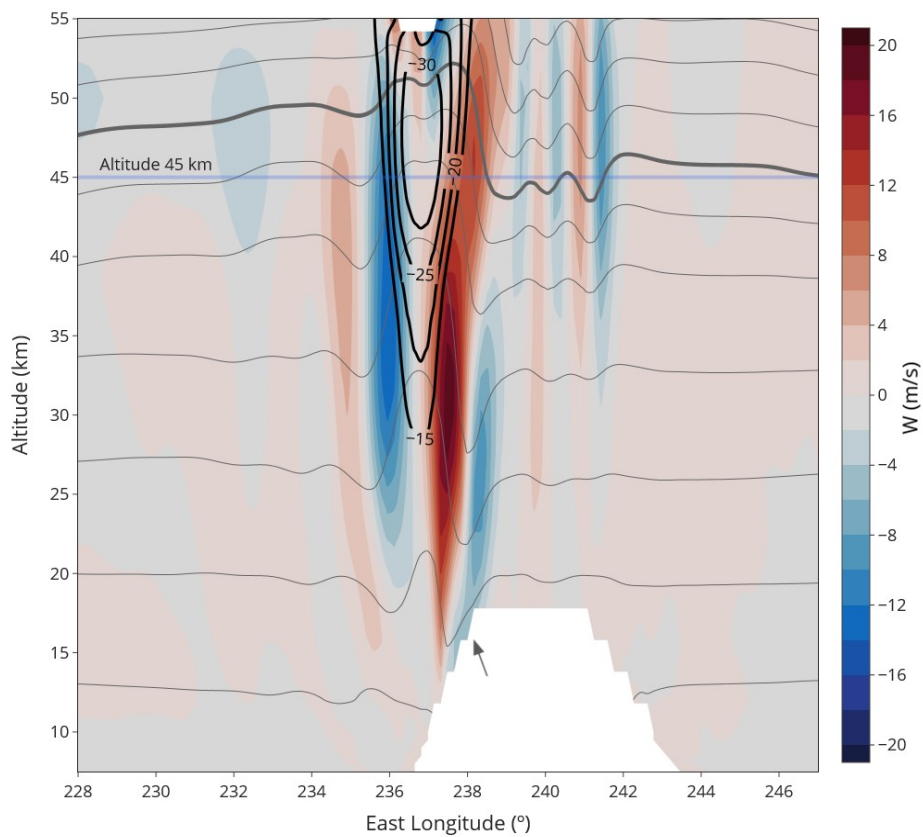

**Abstract**
In a previous work (Hernández-Bernal et al. 2021) we performed an observational analysis of the Arsia Mons Elongated Cloud (AMEC), which stands out due to its impressive size and shape, quick dynamics, and the fact that it happens during the martian dusty season. Observations show that its morphology can be split in a head, on the western slope of the volcano of around 120 km in diameter; and a tail, that expands to the west reaching more than 1000 km in length, making the AMEC the longest orographic cloud observed so far in the solar system. In this work we run the LMD (Laboratoire de Météorologie Dynamique) Mesoscale Model to gain insight into the physics of the AMEC. We note that it is coincident in terms of local time and seasonality with the fastest winds on the summit of Arsia Mons. A downslope windstorm on the western slope is followed by a hydraulic-like jump triggering a strong vertical updraft that propagates upwards in the atmosphere, causing a drop in temperatures of down to 30K at 40-50 km in altitude, spatially and temporarily coincident with the observed head of the AMEC. However the model does not reproduce the microphysics of this cloud: the optical depth is too low and the expansion of the tail does not happen in the model. The observed diurnal cycle is correctly captured by the model for the head of the cloud. This work raises new questions that will guide future observations of the AMEC.

**Plain Language Summary**
This is the second paper of our research on the Arsia Mons Elongated Cloud (AMEC), which is a visually impressive cloud on Mars. It appears on the western flank of the Arsia Mons volcano during a specific season right at sunrise. For three hours it grows, developing a thin elongated tail that has been observed to be as long as 1800 km. In our previous work we described available observations. In this work we run a high resolution atmospheric model that captures the effect of the Arsia Mons volcano on the atmosphere. This model shows that due to the presence of the volcano and its effect on the wind, air is forced upwards next to the volcano, leading to a drop in temperatures of 30ºC, which causes the formation of the cloud under extreme conditions of humidity. This is a success of the model that provides a new understanding of this outstanding cloud, however, the accurate physics behind the extreme expansion of the AMEC are not fully understood yet. This work solves some questions and raises many new ones, which will be an aid in the planning of new observations.

## 1. Introduction

In a recent work (Hernández-Bernal et al., 2021; henceforth paper I) we reported the existence of the Arsia Mons Elongated Cloud (AMEC), and we performed a comprehensive observational analysis of this unusual cloud. It forms every morning on the western slope of the Arsia Mons volcano on Mars, in a not well constrained period within the dusty season, potentially lasting from Ls (Solar Longitude) 220º to Ls 320º, which includes the perihelion (Ls 251º) and the southern solstice (Ls 270º). This orographic cloud starts its expansion from the western slope of Arsia Mons around local sunrise (5.7 local time), and it grows to the west for 2-3 hours. The expansion velocity was observed to be around 130m/s in most Martian Years (MY), and 170m/s in MY34. Its maximum observed length was 1800 km. The altitude of the cloud was measured to be around 45 km. Fig. 1 is a visual summary of paper I featuring its main characteristics.

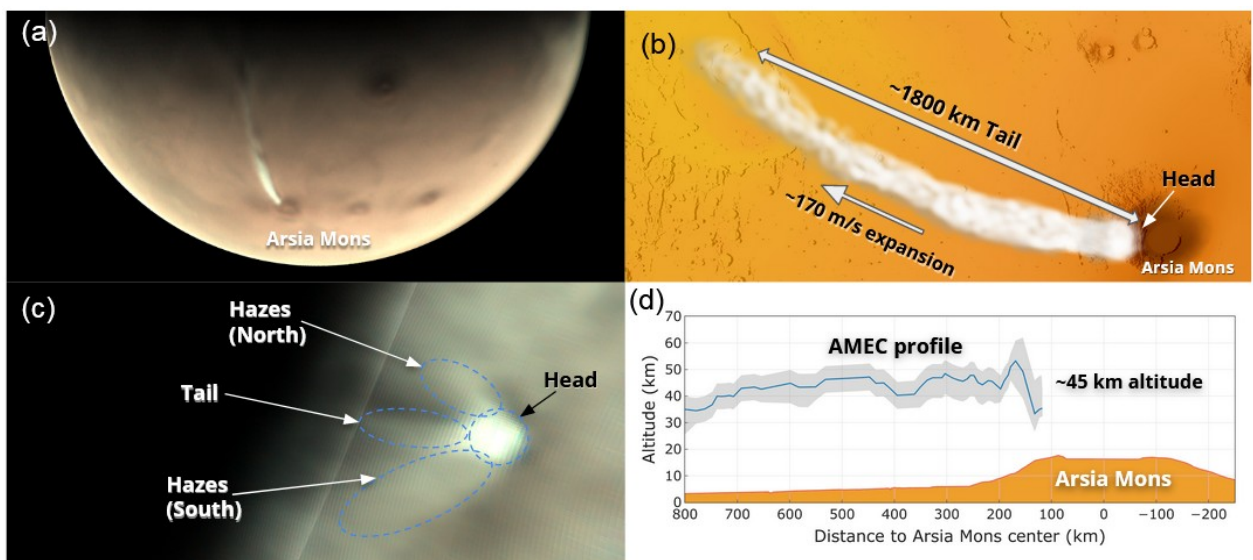

Figure 1. Visual summary of the main AMEC features as reported in paper I. (a) The AMEC as imaged by VMC onboard Mars Express. (b) Idealized artistic impression of the AMEC featuring its main morphological parts (head and tail), its expansion velocity, and tail length. (c) Annotated image of the AMEC (extracted from panel 9g from paper I) showing the hazes occasionally observed around the tail. (d) Vertical profile of the AMEC as reported in paper I.

Arsia Mons is part of the volcanic region of Tharsis, where other volcanoes are also present (see for example Fig. 2). This area has been known for the frequent presence of thick clouds since the early telescopic observations (Slipher, 1927; Smith & Smith 1972). Later on, new observations made obvious that Tharsis is an elevated area hosting high topographic features and that those clouds were of orographic origin (Sagan et al., 1971; Peale et al., 1973).

The spatial and seasonal distribution of water ice clouds through the Martian Year, and the seasonal trends of orographic clouds on the volcanoes of Tharsis were clearly described by Wang and Ingersol (2002), and Benson et al. (2003; 2006) based on daily observations during the local afternoon by the Mars Orbiter Camera (MOC). They showed that orographic clouds are regularly present at the Tharsis volcanoes during the first half of the Martian year (Ls 0º-180º), when tropical clouds are very common as part of the Aphelion Cloud Belt (ACB; Clancy et al., 1996). However, during the dusty season at the second half of the year (Ls 180º-360º), clouds are not common in low latitudes. Systematic observations usually take place in the afternoon, and Arsia Mons is the only low-latitude location with frequent clouds in these observations. The AMEC occurs at Arsia Mons during this season in the early morning, a local time not covered by MOC and other instruments and therefore it is not included in the aforementioned works (see paper I).

Several authors have used Mesoscale Models (MM) to study orographic clouds on Tharsis during the first half of the year (Ls ~100º). Michaels et al. (2006) explored the formation of orographic clouds on Olympus and Ascraeus Mons based on the MRAMS mesoscale model (Rafkin et al. 2001). They showed that the main mechanism behind these orographic clouds is the pumping of water from lower to higher altitudes by insolation driven wind on the slopes of the volcanoes. Water vapor is elevated to higher layers, and it condenses into water-ice clouds that are usually advected to the west by the predominant winds. Spiga and Forget (2009) tested the LMD (Laboratoire de Meteorologie Dinamique) Mars Mesoscale Model on different meteorological phenomena, including orographic clouds on Tharsis at Ls 90º and 120º, and they reached similar conclusions to those of Michaels et al. (2006).

Due to the proximity to the aphelion (Ls 71º), in the current climate global average temperatures on Mars are lower during the first half of the year than during the second half. As a result, during that season the low-latitude hygropause (the layer where conditions for water condensation are met, typically confining water beneath) is located at a relatively low altitude (about 10-20 km), favoring the occurrence of the ACB (Clancy et al., 1996; Montmessin et al. 2004). During the second half of the year, the hygropause raises to 30-60 km and as a result, clouds form at higher altitudes (Clancy et al. 1996), and water vapor is no longer confined to the lower atmosphere (e.g. Fedorova et al. 2021).

In this work we used a hierarchy of multiscale modeling, from global climate simulations to regional "mesoscale" simulations of the area of Tharsis during the second half of the year, and gained insight into the atmospheric physics driving the AMEC. We also use this modeling framework to put the AMEC in perspective with the orographic cloud formation mechanism described by Michaels et al. (2006) and the relationship to seasonal variability of the hygropause.

In section 2 we describe our simulations. We present our results for the AMEC analysis in sections 3, 4, 5 and 6. In section 3 we overview the environment conditions as extracted from the global-scale model. In section 4 we show the prediction of the mesoscale model for the AMEC, and describe the general dynamics and microphysics that very probably lead to the occurrence of this unusual cloud. In section 5 we show the diurnal cycle of the AMEC as predicted by the model. In section 6 we explore sol-to-sol, seasonal, and interannual variations based on the GCM predictions from section 3 and the knowledge about the AMEC acquired in section 4.  In section 7 we provide some conclusions and perspectives.

## 2. Methodology

In this study, we use the Mars Mesoscale Model (MMM; or MM, for Mesoscale Model) initially described by Spiga and Forget (2009), which consists in coupling an adapted version of the terrestrial Weather Research and Forecast model (Skamarock et al. 2008) with the Martian physical packages developed at Laboratoire de Météorologie Dynamique (LMD): radiative transfer (Forget et al. 1999), dust processes (Madeleine et al. 2011), and water-ice clouds formation including microphysics (Navarro et al. 2014). The MMM uses initial and boundary conditions from the LMD Global Climate Model (GCM) with which most of the physical parameterizations are shared. The mesoscale modeling methodology deployed in this paper is similar to mesoscale simulations developed in Spiga et al. (2017).

Specifics of the simulations carried out to explore AMEC-like phenomena are detailed as follows. The horizontal domain comprises 321 x 321 grid points with mesh spacing of 10 km, encompassing the large region of the Tharsis volcanoes around Arsia Mons (see Figure 2). Along the vertical, 101 levels were included with a model top at 1 Pa (approximately 60 km above the local surface) and mesh spacing progressively increasing from 5m in the lowermost layer to 1300m at model top; vertical resolution within the altitudes of interest for this study (20 km and 45 km) are respectively about 700m and 900m. Integration timestep is 20s and the WRF dynamical core uses the non-hydrostatic option. All simulations were run for three full Martian sols; as is usually done for mesoscale simulations, the first sol is considered as model spin-up and the second and third sols are retained for analysis. Results from the second and third simulated sols are very similar and we chose to perform our analysis on the second simulated sol. Plots refer to this second sol, except otherwise indicated.

Our reference simulation to analyze AMEC-like signatures is constrained by the dust scenario for MY34 detailed in Montabone et al. (2020). We chose this scenario because most observations in paper I were in MY34. Note that the GCM predictions for this MY34, which were used for mesoscale simulations as initial and boundary conditions, are also discussed in our section 3 to discuss the global-scale conditions relevant for our analysis. To put MY34 in a broader context, this discussion in section 3 also includes GCM runs using dust scenarios at other Martian Years (Montabone et al. 2015). The 3-sol MMM simulation is done at the northern winter solstice at Ls=270°. In this paper, all local times (LT) are Local True Solar Time at the meridian 240°E, which crosses Arsia Mons and all altitudes are expressed as being above the reference areoid defined with MOLA topography (Smith et al., 2001).

The LMD MMM predicts the observed daytime and nighttime cloud covers for the aphelion season around Ls=120° (Spiga and Forget 2009, their Figure 10; Spiga et al. 2017, their supplementary figure 5), similarly to predictions discussed e.g. in Michaels et al. 2006 (their figure 1). Previous reports of mesoscale modeling of water-ice clouds for Tharsis did not discuss the perihelion season Ls=270° considered here. The AMEC takes place in the morning, when not many observations are available, probably all of them analyzed in paper I. On the other hand, systematic observations in the afternoon by sun-synchronous orbiters have enabled a very good knowledge of that local time. Based on such observations, Wang and Ingersol (2002) and Benson et al. (2003; 2006) showed that during the perihelion season Arsia Mons is the only Tharsis volcano that exhibits prominent clouds in the afternoon, which is in qualitative agreement with the MM results shown in Figure 2 for the same local time. This strengthens the reliability of our results.

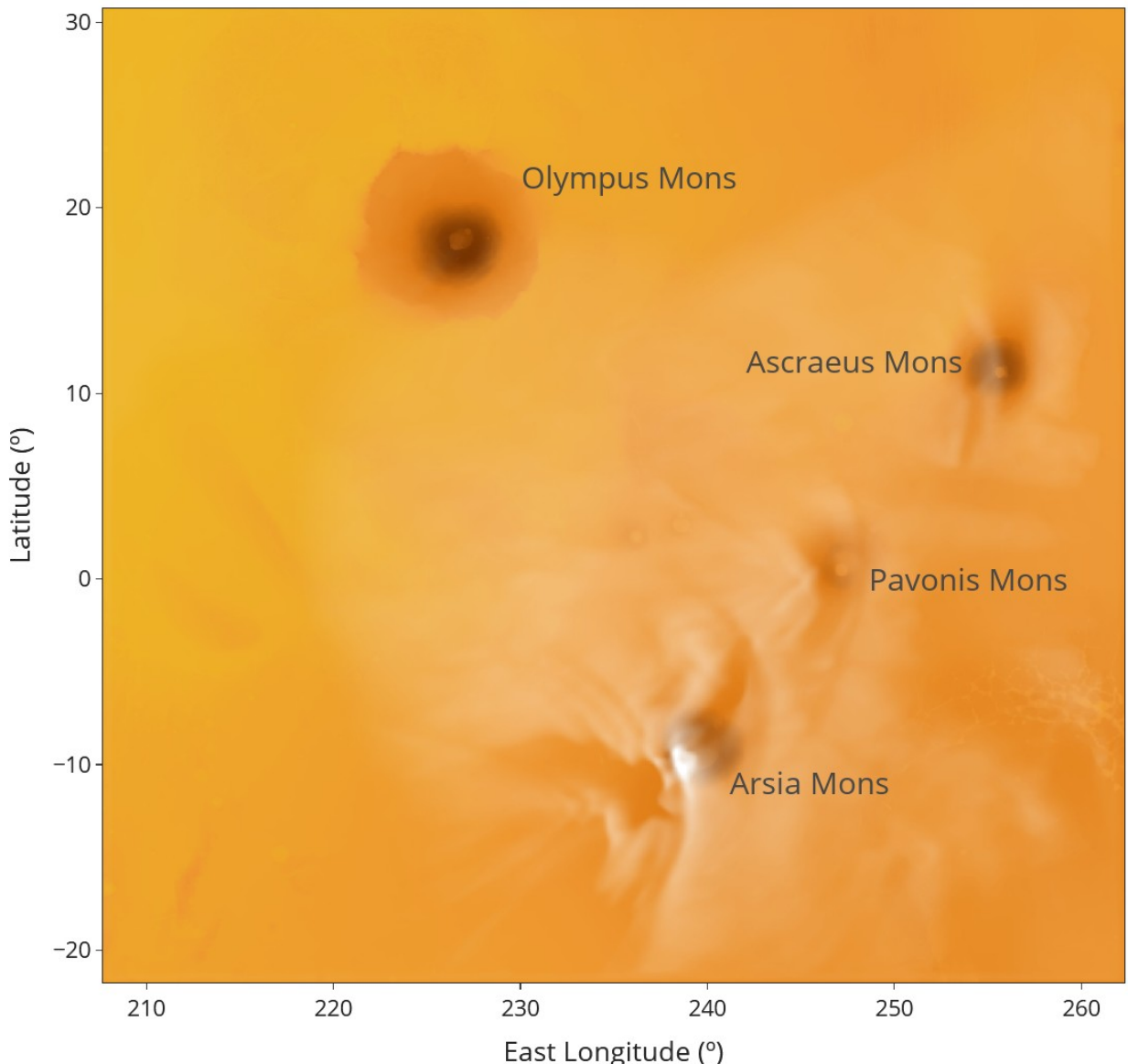

Figure 2. Representation of visible clouds predicted by the MM in the afternoon in the reference simulation, in MY34 at Ls 270º. The maximum predicted optical depth is around 0.2. Refer to supporting figure S1 for the equivalent quantitative representation of optical depth.

## 3. Environment conditions from the GCM

We show in this section the relevant unperturbed environment conditions at the time and season of the AMEC, as extracted from the GCM. We expect winds to be a key for the dynamics of the AMEC, and temperature and water content to affect the microphysics of the cloud. These large scale variables are given by the global circulation, which also involves the relevant effects of thermal tides (e.g. Barnes et al., 2017). Since the AMEC occurs at a particular season and local time, we show these variables in terms of altitude, local time, and solar longitude.

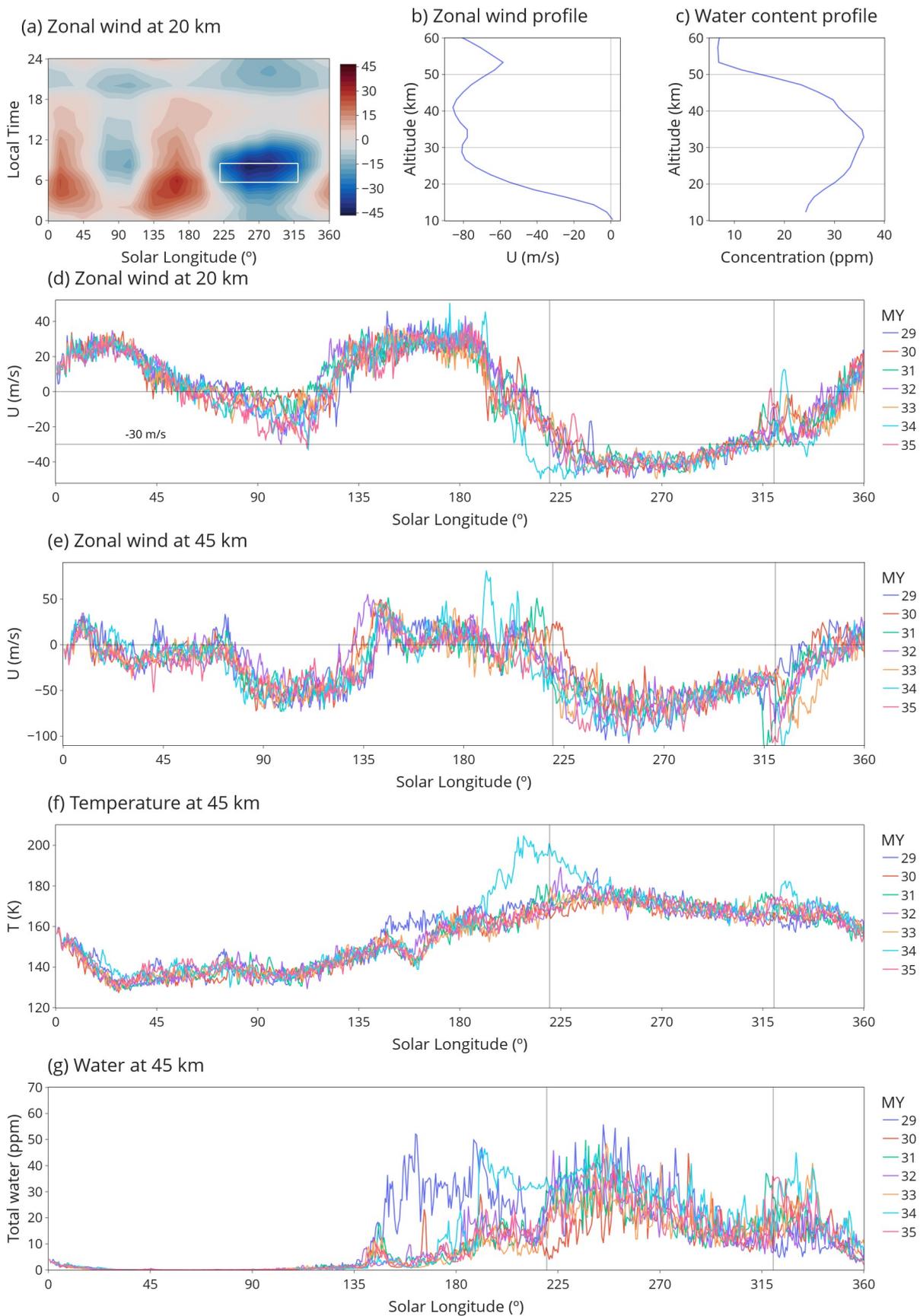

Figure 3. Environment conditions. Negative zonal winds are westward. (a) Zonal winds (m/s) over the summit of Arsia Mons (at 20km), extracted from the Mars Climate Database. The strongest winds are coincident with observations of the AMEC: at Ls ~220º-320º, and Local Time 5.7-8.5 (squared area). (b) Vertical profile of zonal winds on Arsia at 7LT and Ls 270º. (c) Vertical profile of water content (vapor and ice) on Arsia at 7LT and Ls 270º. (d-g) Variables at relevant altitudes for each sol of different years in the GCM, at 7 LT at Arsia Mons; seasonal

trends, sol-to-sol variations, and interannual variations are apparent. Potential AMEC season indicated between two vertical lines. (d) Zonal winds at 20 km (over the summit of Arsia). (e) Zonal wind at 45 km (cloud altitude). (f) Temperature. (g) Total water (vapor and ice) volume mixing ratio (ppm).

Figure 3a shows zonal winds at 20 km (over the summit of Arsia Mons) plotted in terms of Solar Longitude and Local Time for MY34. It is apparent that the AMEC season and local time (marked by a white rectangle) is coincident with the fastest zonal winds on the summit of Arsia. This trend repeats annually, although there are obvious interannual variations (see fig. 3d). The early onset of strong westward winds in MY 34 is particularly noticeable. Vertical zonal wind profiles show westward winds higher up, peaking at approximately 45km (Fig. 3b shows the profile at 7LT and Ls270, when the cloud is fully developed), and winds at 45 km are also faster during the AMEC season compared to other seasons (fig. 3d).

As expected, temperature is higher during the perihelion season, due to the higher insolation (fig. 3f). This leads to a rise of the hygropause, which enables higher concentrations of water in the mesosphere (fig. 3c and 3g).

## 4. Dynamics and Microphysics of the AMEC
In this section, we use results from the mesoscale model (MM) to identify the main physical processes that lead to the occurrence of the AMEC.

As shown in paper I, the AMEC occurs in the season around Ls 220-320º, and its daily expansion starts around 5.7 LT, finishing around 8.5 LT. For simplicity, in order to explore the general mechanisms of the phenomenon, we use in this section a fixed local time, 7.6LT, a local time when the AMEC is well developed in observations. The diurnal cycle is explored in section 5.

Water pumping by volcanoes from lower altitudes (by upslope anabatic winds) is an efficient mechanism to form orographic clouds at low latitudes during the aphelion season (Michaels et al., 2006). Our simulations show that this is not the main mechanism behind the AMEC (or other orographic clouds taking place on Tharsis during the perihelion season), because the hygropause moves to higher altitudes during this season compared to the colder aphelion season (e.g. Fedorova et al., 2021). As a result, at the perihelion season, less water is available at low altitudes to be transported upwards by volcanoes (see for example figs. 3c and 6d), and the altitudes reached by transported water are not high enough to trigger condensation. Instead, gravity waves produced by the volcanoes, which perturb the hygropause, are the main driver of orographic clouds during the perihelion season. This is illustrated in supporting videos S1 and S2. Taking this into consideration, we base our following discussion about the AMEC on the fact that the vertical transport of water from the surface is negligible.

### *4.1. Perturbation of the atmosphere by Arsia Mons during the AMEC*
The MM shows how Arsia Mons perturbs the local atmosphere at all altitudes during the AMEC season (Fig 4). Easterly winds are perturbed by the volcano, triggering orographic gravity waves, creating ascending and descending cells, with the strongest updraft on lee side (red cell in the left side of the volcano in Fig. 4), at the western side of the volcano. Typical gravity wave perturbations of vertical wind simulated by the model reach +/- 8 m/s, while the strongest updraft reach a much larger vertical velocity of 20 m/s with no corresponding downdraft of this amplitude. This Strong Vertical Updraft (SVU from now onwards) causes a drop in temperatures (a cold pocket; represented by thick black contours in Fig. 4) of down to -30K at altitudes around 40-50 km. The longitude, latitude and altitude of the center of this cold pocket are coincident with the position of what we called the head of the AMEC in paper I (see fig. 1). The PFS instrument onboard Mars Express has recently observed cold pockets over Tharsis volcanoes in this season that resemble the one found in our simulation (P. Wolkenberg, private communication)

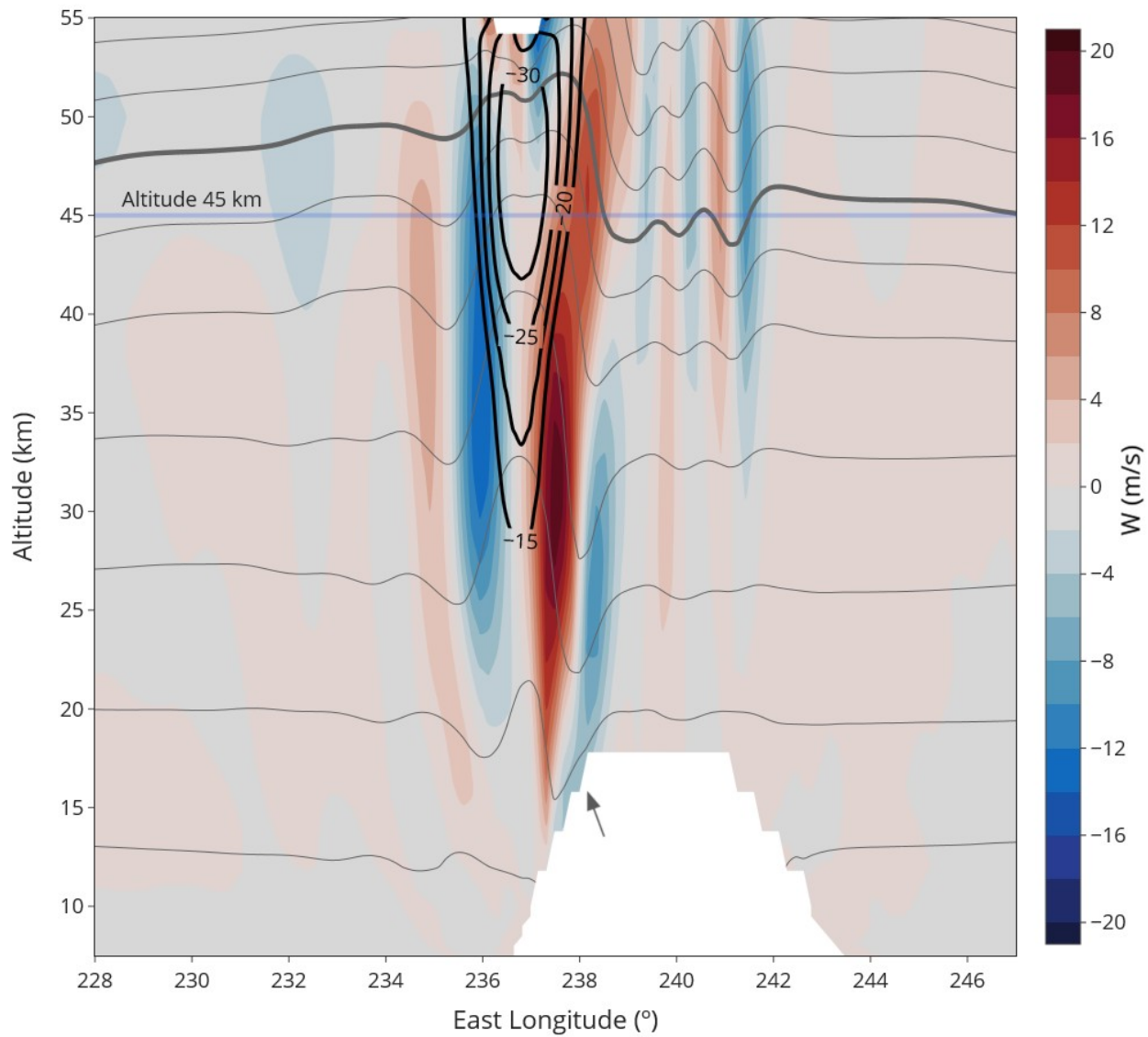

Figure 4. Atmospheric perturbations caused by Arsia Mons in the reference run of the mesoscale model at 7.6 LT at latitude 8.7ºS. Background shade represents vertical winds in m/s (scale at right). Gray curves are potential temperature contours, which correspond to streamlines, with zonal winds blowing westward (i.e. from right to left in this graph). The thick gray curve departing at 45 km shows that there is a positive vertical displacement after the head of the cloud. Thick black contours indicate the area where the temperature is lower than the mean for a given altitude, from -15K to -30K. A horizontal blue line indicates 45km in altitude for comparison with latter figures. The arrow in the mountain indicates the lee side downslope windstorm.

This SVU in the lee side is strongly reminiscent of a hydraulic jump. Upstream the hydraulic-like jump, on the slope of the mountain, we see a downslope windstorm (blue cell on the left slope of the volcano in Fig. 4, indicated with an arrow). Downslope windstorms on Earth mountains have received strong interest because they can be very fast near the surface, potentially posing a risk, and they are sometimes followed a hydraulic jump. Different theoretical approaches have been proposed to characterize this kind of flow around mountains (e.g. Durran 1990; Lott, 2016). In our case, the fast winds blowing on the summit of Arsia Mons (Fig. 3a), the vertical shear of winds, and the presence of a critical level near the surface (Fig. 3b), are conditions that favor the occurrence of the downslope windstorm followed by the hydraulic-like jump.

The SVU grows with altitude (in terms of wind speed and flow perturbation) as a result of the lowering of pressure, as expected from gravity-wave theory. The adiabatic ascent of air masses in the SVU is the cause of the temperature drop, which is larger for higher vertical ascent of air parcels. The thick gray streamline shown in fig. 4, reveals a net ascent of up to 8 km at the altitude where the AMEC forms (~45km), coincident with the largest drop in temperature. It is followed by much slower descent that retains colder temperatures in the whole area, as we show in section 4.2. The total vertical ascent of parcels is not just a result of vertical wind speed, the width of the ascending cells, and the horizontal wind speeds are also important.

As we noted in section 3, the AMEC season coincides with the fastest zonal winds on the summit of Arsia Mons (Fig. 3a). We cannot establish a definitive connection between these transient fast winds and the occurrence of the SVU leading to the AMEC. Other variables, like the specific vertical profile of winds or potential temperature might play a significant role.

In summary of this section, the occurrence of the AMEC is coincident with the presence of fastest winds on the summit of Arsia Mons (Fig. 3a). This, together with the characteristics of the wind profile is probably the cause of the onset of the SVU on the lee of the volcano. At higher altitudes, at the altitude of the cloud, the adiabatic ascent of air parcels for up to 8 km creates a cold pocket with temperatures down to 30K below the environment temperature (Fig. 4), coincident in space and time with what we called the head of the AMEC in paper I.

### *4.2. Anatomy of the Cold Pocket*
In fig. 4, thick contour lines mark the position of the cold pocket corresponding to the head of the AMEC at 8.7ºS. The perturbations caused by Arsia Mons in the atmosphere extend in latitude in an asymmetric way that we describe in this subsection. Fig. 5 shows the extent of the cold pocket at 45 km. The environment temperature at this altitude is around 160K in the MM.

Next to the SVU, where the ascent of air parcels as traced by streamlines in fig. 4 is maximum, the cold pocket reaches its lowest temperature, below 140K (ΔT=-20K), and down to 130K (ΔT=-30K). This lowest-temperature area has a round shape that visually resembles the aspect of the head of the AMEC in observations. Its diameter is around 120km, its center latitude is 8.7ºS, and its easternmost longitude is 237.5ºE; all these parameters are in agreement with the measurements shown in figure 8 of paper I, and therefore we consider that this feature predicted by the model corresponds to the actual head of the AMEC in observations.

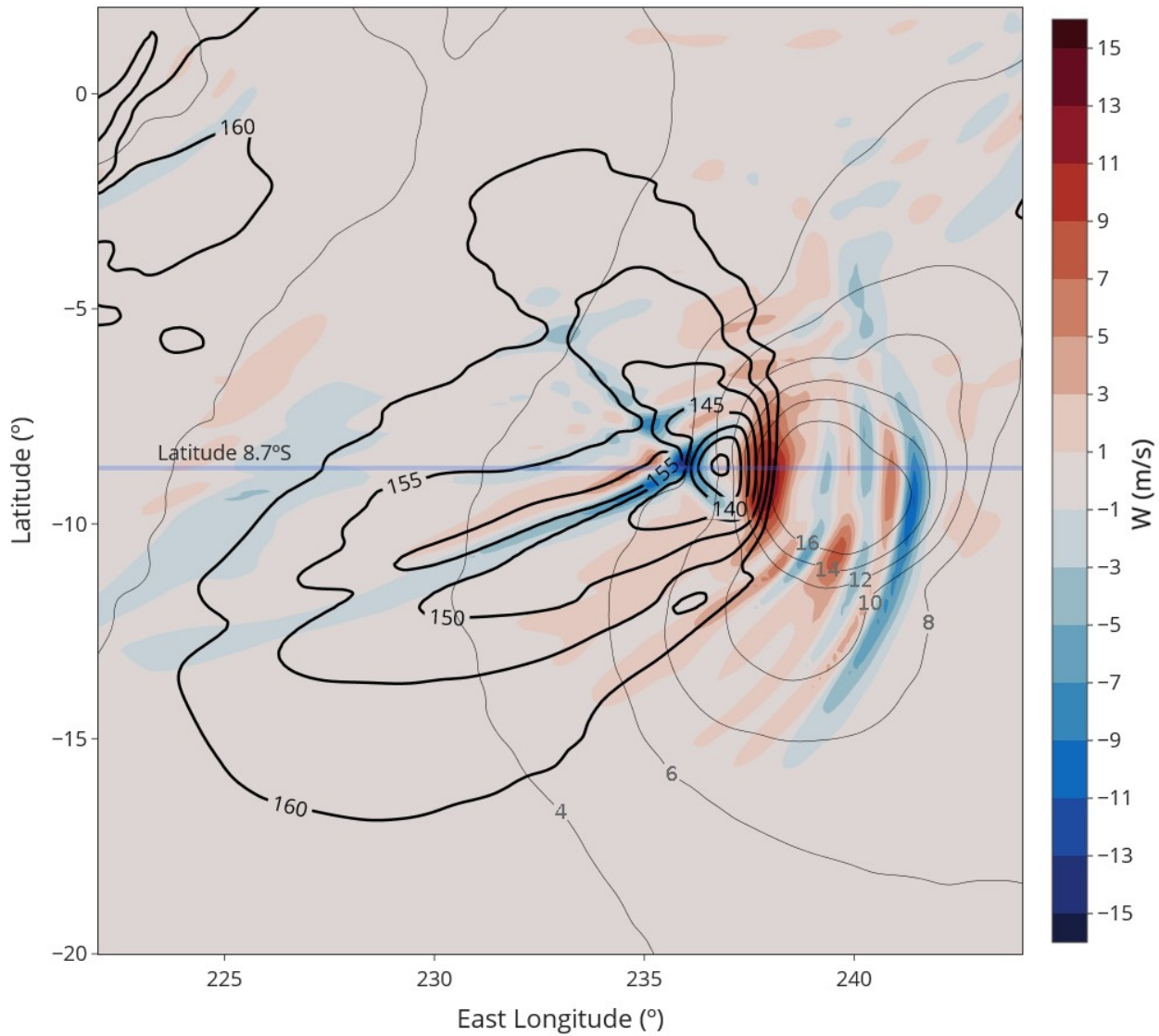

Figure 5. Anatomy of the cold pocket at 45 km in altitude. Background shade shows vertical winds in m/s, thin grey curves are contours of topography (altitude indicated in grey numbers), bold contours are isotherms below 160K, which is around the environment temperature. The core of the cold pocket (T=130-140K) is coincident with the head of the AMEC, and is located right over the SVU. From this position, two branches with temperatures below 155K depart to the northwest and southwest. A bigger area below 160K extends to the west for almost 1000 km. A horizontal blue line indicates latitude 8.7ºS for comparison with figure 4.

The contour below 155K (ΔT<-5K) exhibits the shape of two branches extending to the northwest and southwest. The southern branch has a narrow inner region of lower temperatures T<150K that extends for around 500 km, with a width of only around 100 km. It is tempting to consider that this branch of the cold pocket corresponds to the tail of the AMEC, but we will show in section 4.3 that the tail is more likely caused by advection. Both branches of the cold pocket extend to the West, which is the approximate direction of horizontal winds at this altitude. It is likely that the shape of this cold pocket with its two branches is related to the U shaped patterns found in the theoretical work by Smith (1980) on the flow past an isolated mountain.

Finally, the area where the temperature is below the unperturbed 160K (ΔT<0K), extends up to 1000 km to the West, due to the slow descent of air parcels after the SVU described above.

### 4.3. Cloud formation and expansion of the AMEC

In this subsection, we explore the cloud formation in the head of the AMEC and the formation of the characteristic tail.

We start by analyzing the water vapor and ice distributions in the model. Fig. 6a and 6b show water ice distribution and total amount of water respectively, at 45 km altitude. The total amount of water is increased in the head of the AMEC compared to other areas at the same altitude. The vertical distribution of total water (Fig 6d) shows that this is a result of the vertical displacement of moist parcels following the SVU, which transports water from lower layers of the atmosphere, where water concentration is higher.

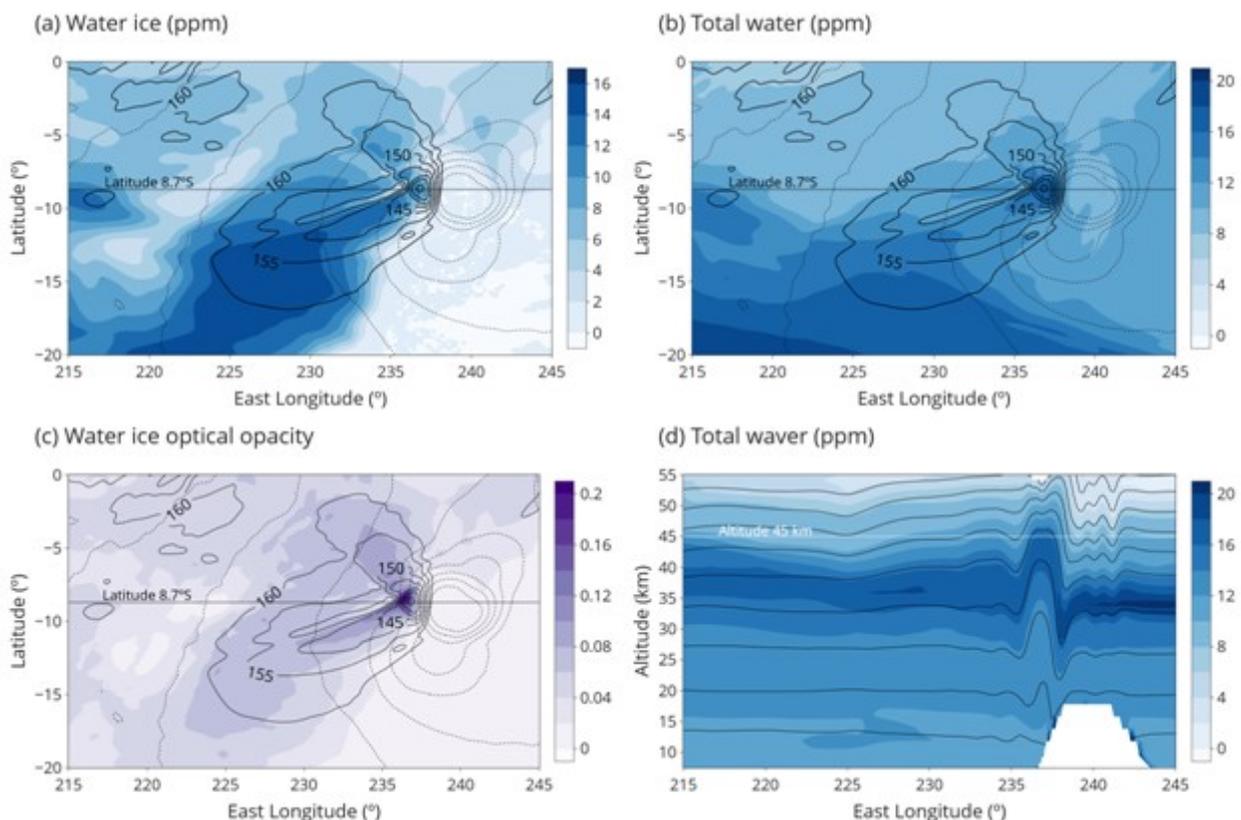

Figure 6. Water and clouds around Arsia Mons during the AMEC at 7.6 LT as predicted by the mesoscale model. Color scales represent the magnitude indicated at the top of each panel. In panels a-c thick contour lines are isotherms, and in panel d lines of constant potential temperature, which correspond to streamlines. (a) Water ice concentration at 45 km. (b) Total water concentration at 45 km. (c) Water ice optical depth. (d) Total water volume mixing ratio (ppm) at latitude 8.7ºS.

Figs. 6a and 6c show that water ice and optical depth are higher in the position of the head. However, the model does not directly reproduce the observed tail of the AMEC. Additionally, the maximum optical depth at the core of the cold pocket is 0.2, a low value, considering the brightness of the AMEC in the images by different instruments shown in paper I.

In the following subsections we discuss the microphysics of cloud formation in the head (section 4.3.1.), the microphysics that might enable the existence of the tail (section 4.3.2.), and the sensibility of the microphysics with the contact parameter (section 4.3.3.).

### *4.3.1. Microphysics of the head*
Fig. 7 shows the relative humidity index at 45km in logarithmic scale. Water vapor in excess of saturation exists wherever this index is higher than 1 (0 in logarithmic scale), therefore this figure reveals that a large amount of water remains in excess of saturation around the head of the AMEC in the model. The area where ΔT<-20K, which is the area that we identify as the head of the AMEC, is characterized by a relative humidity index over 200 (2.3 in logarithmic scale). The relative humidity index reaches values as high as 5000 in the core of this area (3.7 in logarithmic scale).

Water vapor in excess of saturation was first observed on Mars by the SPICAM (Spectroscopy for Investigation of Characteristics of the Atmosphere of Mars) spectrometer onboard Mars Express (Maltagliati et al. 2011). Early models of the Martian atmosphere considered a simplified cloud microphysics that assumed that all water vapor in excess of saturation condenses into water ice, and thus could not account for supersaturated vapor. Microphysics of LMD Mars models (including the mesoscale model used in this work) was improved by Navarro et al. (2014). Among other things, they considered nucleation processes, allowing condensation only when suitable condensation nuclei are present and therefore enabling the model to predict the presence of water vapor in excess of saturation. Indeed, Navarro et al. found in their GCM runs supersaturation indexes in the order of ~5000 (their figure 10).

Cloud particles generated within the core of the cold pocket must be small, because they only stay for a few minutes in the coldest area where ΔT=-20K. The zonal wind speed at 45km is ~100m/s, so an air parcel crosses this coldest region, of ~100 km, in ~10$^3$s, while the characteristic timescale for particle growth by condensation is in the order of 10$^4$s (Clancy et al. 2017, their figure 5.18). We do not count with measurements of particle size for the AMEC, however, the cloud trails reported by Clancy et al. (2009; 2021), which share many characteristics with the AMEC, exhibit particle sizes below 1μm.

If cloud particles are small, and so much water vapor remains in excess of saturation, the optical depth will be very sensitive to the availability of condensation nuclei. Thus an enhanced concentration of condensation nuclei (dust particles) in the head of the AMEC, might be able to reduce the water vapor in excess of saturation and increase the cloud optical depth in the model.

### *4.3.2. Microphysics and expansion of the tail*
There are at least two different mechanisms that could be involved in the expansion of the AMEC: Advection and in-situ condensation.

The morphology of the southern branch of the cold pocket shown in the previous subsection is reminiscent of a long tail, and suggests that in-situ condensation in the southern branch could be the origin of the tail, and advection would not play a significant role. However, there are caveats with this interpretation. Both northern and southern branches of the cold pocket are associated with higher values of water ice and optical depth (figs. 6a, 6c). If the tail corresponded to the southern branch, we would expect another cloud coincident with the northern branch of the cold pocket that is not observed. In our interpretation of the model, this optically thin cloud corresponds to the observed hazes around the AMEC, which occasionally appeared forming almost symmetrical angles with respect to the tail (see fig. 1c).

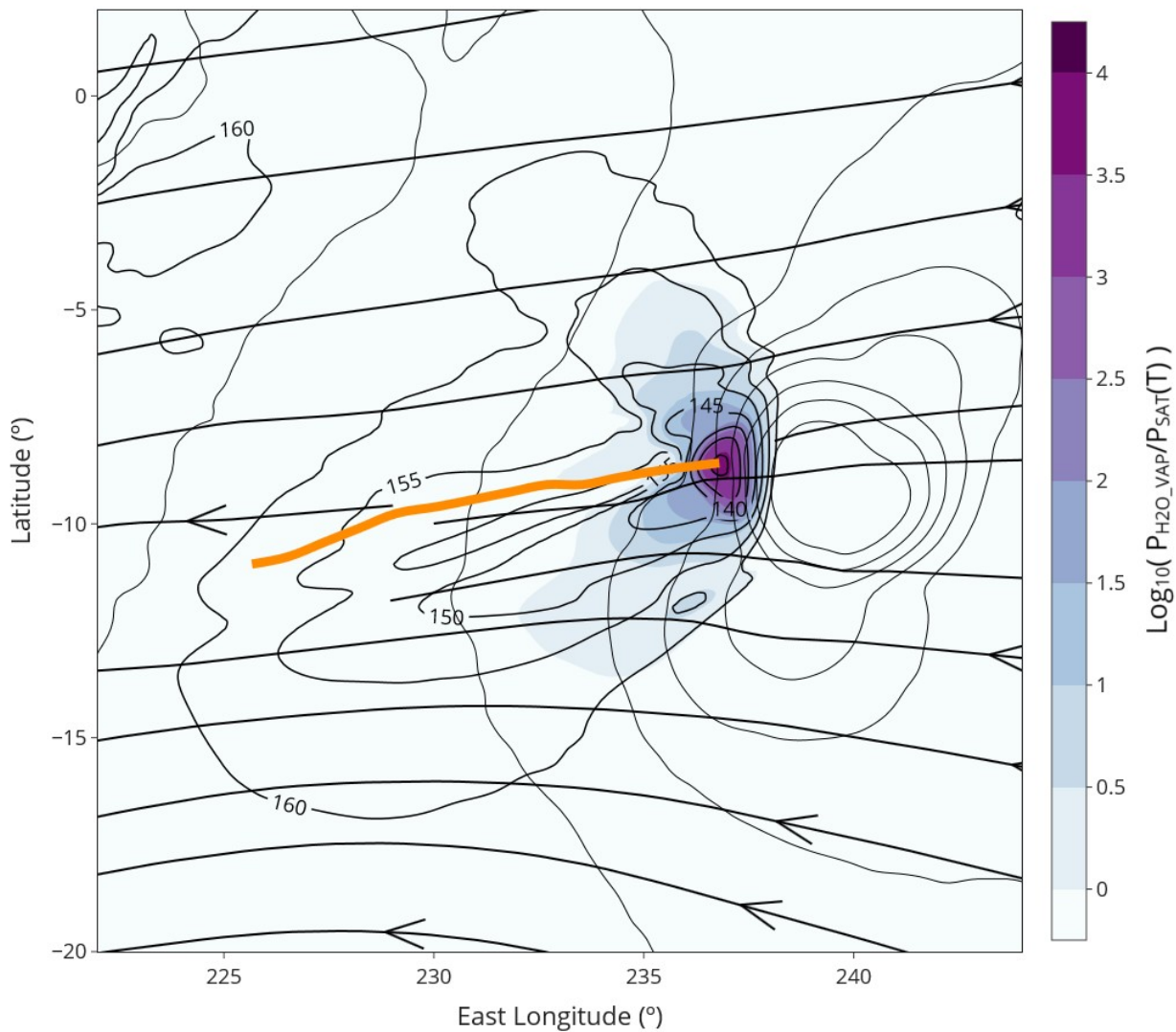

Figure 7. Relative humidity index at 45 km: water vapor partial pressure divided by the saturation pressure at a given temperature: PH2O_VAP/PSAT(T). Represented in log10 scale, as extracted from the mesoscale model at 7.6 LT. Superimposed streamlines of winds (arrows), and advection trajectory traced by hypothetical cloud particles from 5.7 to 7.6 LT (thick orange curve departing from the head). Cold pocket in the same format as in previous figures for reference.

Another possibility, as we proposed in paper I, is that condensation occurs mostly in the core (the head of the AMEC), where the cold pocket reaches its lowest temperatures. The core of the cold pocket coincides in location and extension with the observed head of the AMEC, and the narrow tail can be explained as the result of the advection of cloud particles that condense in this small region due to increased relative humidity.

We expect that cloud particles condense in high numbers in the head of the AMEC, due to the extreme saturation conditions found there. These particles would then be advected following the trajectory indicated schematically by the orange curve in fig. 7. In the model, condensed particles sublimate as they leave the core of the cloud, and the model does not predict an important optical depth in the advection trajectory. The fact that the advection trajectory traverses the extended cold pocket, where temperatures are still low, could contribute to a slower sublimation, that would favor the AMEC tail.

The advection trajectory shown in fig. 7 reaches a length of around 600km in 1.9h, which corresponds to a mean velocity of ~90m/s. In comparison, we reported in paper I a velocity of 170±10 m/s in MY34, and 130±20m/s in other years. This large value of the expansion velocity repeated consistently from sol to sol in MY34 (see fig. 6 in paper I) and is not reproduced by the MM, probably a consequence of an underestimation of background velocities in the GCM (see figure 3d). In addition, the GCM (Fig. 3d) predicts sol-to-sol variations of around 20-30 m/s that were not observed in MY34 (there were not enough observations in previous years).

### 4.3.3. Sensibility to contact parameter
Microphysics of the model do not reproduce the optical depth of the head or the advected tail of the AMEC. We tried to test microphysics by changing the contact parameter, which is a parameter related to the activation of condensation nuclei (see the nucleation theory described by Montmessin et al., 2022). This kind of test was also performed by Navarro et al. (2014).

The default contact parameter in the reference simulation was 0.95, and in addition, we tested 0.85 and 0.99. Results are presented in supporting figure S2. The optical depth of the head does not change significantly, and the tail of the AMEC is still not reproduced in any case. Differences in the optical depth are present mostly in the branches of the cold pocket.

The result of this test indicates that variations in the contact parameter do not improve the reproduction of the AMEC in the model.

---

In summary of this subsection. Extreme condensation conditions are present in the head of the AMEC as predicted by our simulation, however, it predicts low optical depth in the head compared to observations, and the tail of the AMEC is not reproduced. The model is not capturing with accuracy the microphysical processes involved in the AMEC, maybe because of a lack of condensation nuclei, and we propose that due to the extreme conditions in the head of the AMEC, a large number of small water ice particles could form there, and be advected by easterly winds, resisting sublimation both because they are small and because they traverse a cold area. Predicted winds do not match the observed expansion of the tail.

**5. Diurnal cycle in the MM**

In section 4, we discussed the physics of the AMEC at a fixed local time (7.6 LT), when the head of the cloud is fully developed. In this section, we explore the diurnal cycle of the AMEC as revealed by the mesoscale model, and compare it with observations. Supporting videos S3 and S4 show the evolution with local time of figs. 4 and 5 respectively.

During the AMEC season, the onset, expansion and clearing of the cloud repeat daily in a regular way. In paper I we reported that the expansion phase of the AMEC starts with sunrise around 5.7 LT, and finishes around 8.5 LT, as deduced from observations in MY34. After 8.5 LT, the tail detached from Arsia Mons, probably because cloud formation ceased in the head while advection of the tail continued.

The head of the AMEC predicted by the mesoscale model is in good agreement with that diurnal cycle (see supporting video S3). In fig. 8 we illustrate the local-time evolution of different variables related to the strength of the head for the two sols of our reference simulation. As expected, the two sols evolve in a similar way. Zonal winds at 20km (Fig. 8a, see also Fig. 3a), which probably are the ultimate cause of the AMEC, reach their maximum intensity around sunrise. Vertical wind speed at the SVU (fig. 8b) reaches its maximum around 7.5 LT, however, as discussed in section 4.1, the total ascent of air parcels and its consequent temperature drop depend also on the width of ascending cells and on the horizontal wind speed. Maximum drop in temperatures (fig. 8c) and maximum air parcel ascent in the head (fig. 8d) takes place around 8.2 LT.

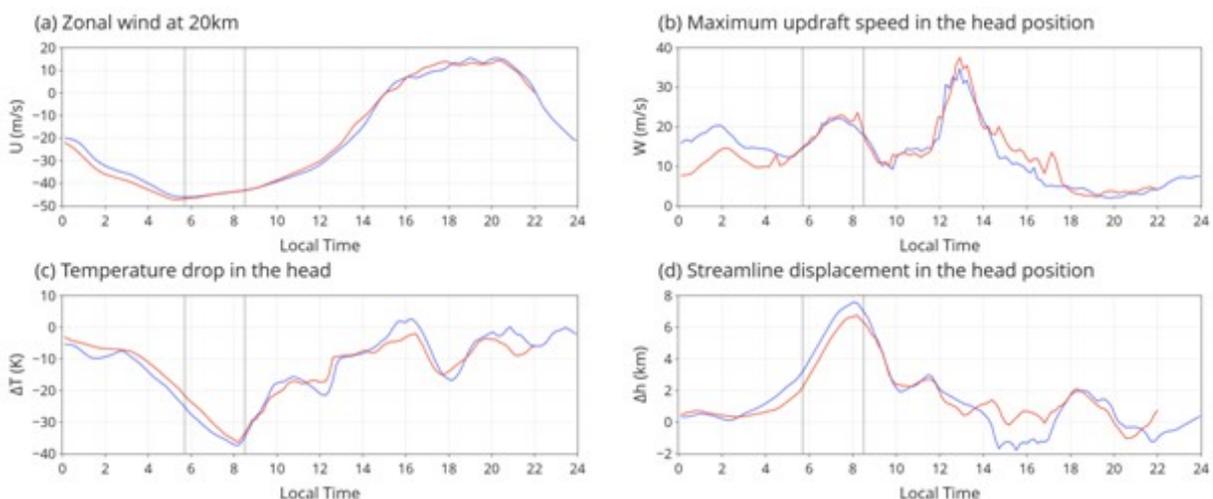

Figure 8. Local Time evolution of selected variables related to the strength of the head of the AMEC. Blue curve corresponds to sol2 in the mesoscale simulation, which is the sol displayed in figures in section 4; red curve corresponds to sol 3. Vertical lines indicate the expected starting (5.7 LT) and end (8.5 LT) of the expansion phase according to observations. (a) Zonal winds at 20km, easterly (negative) winds are strongest around sunrise. (b) Maximum vertical wind speed found in the column below the head. (c) Temperature drop in the head (at 45km). (d) Maximum ascension of air parcels in the head (at 45km), as estimated from the potential temperature field.

As we can see in supporting video S3, the core cold pocket corresponding to the head collapses around 8.5 LT, in very good agreement with the observed end of the head of the AMEC. This might be a result of wave breaking when the amplitude of vertically propagating gravity waves exceeds a threshold.

The evolution of variables presented in in panels b,c,d in fig. 8 is not symmetrical. Their slopes are higher at the end of the AMEC cycle. However, it is not straightforward to set thresholds in those variables that predict the life cycle of the head of the AMEC.

## 6. Sol-to-sol and interannual variations from the GCM

Our previous analysis focused in the results of the mesoscale model for two consecutive sols in the AMEC season in MY34, the year when the behavior of the AMEC is best characterized. In this section we discuss the potential influence of sol-to-sol and interannual variations of the key physical parameters as shown in section 3 from GCM outputs (fig. 3).

Water content suffers strong sol-to-sol variations (fig. 3g). This is likely linked through microphysics to some visual characteristics of the cloud (mostly its brightness), which have been observed to change from sol to sol.

In fig. 3d, we identify a potential threshold for the occurrence the AMEC season: Easterly winds at 20 km are larger than ~30 m/s during that season. Day-to-day variations of environmental winds at 20 km could produce sol-to-sol variations, especially at the starting and end of the season, when wind speeds are close to that threshold. Even if the SVU happens, the tail might not expand. This is what we reported as "early anomalies" (see paper I), observed during the early sols of the AMEC in MY34, Since the head was there, the SVU was probably there too, just the tail did not expand because suitable microphysical conditions at 45 km were not met. Therefore, the starting of the AMEC season can be conditioned by the occurrence of the SVU or by microphysical conditions at 45 km.

We do not appreciate significant interannual variations during the AMEC season in figure 3, with the exception of MY34, a year that was characterized by a Global Dust Storm (GDS) event (Sánchez-Lavega et al. 2019). Fast winds at 20km were present earlier than usual, starting at Ls 207º. However, observations in Ls 223-235º do not show the AMEC, which started in MY34 sometime between Ls 235º and Ls 242º (see paper I). We see in fig. 3e that GCM predicted for MY34 temperatures at 45km 20K higher than usual during the dust storm. The temperature slowly lowers until it reaches values comparable to those predicted by the GCM for other years at Ls 240º - 245º. A higher temperature would prevent cloud formation, and this probably explains the appearance of the AMEC in MY34 after Ls 235º - 243º and not before, even if zonal winds at 20km were already fast.

## 7. Conclusions and perspectives

We have used a mesoscale simulation to gain understanding into the physics behind the AMEC. We see that, due to the rise of the hygropause during the dusty season, the water pumping mechanism described by Michaels et al. (2006) for orographic clouds during the ACB is not the main driver of orographic clouds in this season. Instead, orographically generated gravity waves that reach and perturb the hygropause is the primary mechanism behind orographic clouds. The model predicts a dynamical process triggered by the interaction of winds with Arsia Mons. A downslope windstorm followed by a hydraulic-like jump forms on the western slope of Arsia Mons (the lee slope), which produces a Strong Vertical Updraft (SVU) and leads to the formation of a cold pocket 30K colder than the environment temperature, due to the adiabatic expansion of air cells being transported upwards. The region with the highest drop of temperature ($\Delta T=-20K$) is compact, and spatially and temporally coincident with the observed head of the AMEC, at 45 km in altitude on the western slope of the volcano. The diurnal cycle of the AMEC is correctly reproduced by mesoscale modeling. Transiently fast winds on the summit of Arsia at the local time of the AMEC as part of the general circulation constrains the seasonal cycle.

Cold temperatures in the head imply a drop in saturation vapor pressure that produces cloud formation and extreme values of relative humidity. The model does not reproduce the observed optical depth of the head, but it predicts strong supersaturation. An enhanced availability of condensation nuclei in the model could induce more condensation, reduce the water vapor in excess of saturation and increase the predicted optical depth.

In the model, advected cloud particles sublimate, and the optical depth of the tail is not reproduced. Increased condensation at the head, together with the small size of condensates and the low temperatures in the area (caused by the interaction of the atmosphere with Arsia), which might slow sublimation, could lead to the observed expansion of the tail.

Easterly zonal winds at 45km (Fig. 3e) do not match observations well, and this might have important consequences. The advection trajectory predicted by the MM corresponds to an expansion velocity of ~90m/s, lower than the measured expansion velocity, 170±10 m/s in MY34, and 130±20m/s in other years. This is an important discrepancy.

Our conclusions about the physical mechanism driving the AMEC are partially in agreement with those proposed by Clancy et al. (2021) for the Perihelion Cloud Trails. In their case, the proposed origin of the perturbation leading to the formation of cold pockets was insolation maxima in the afternoon. Since the AMEC happens in the early morning, that was not a possibility in this case. This work demonstrates that wind-surface interaction is another possible origin for this kind of feature.

Most previous works on clouds on Tharsis have focused on the ACB season (i.e. Ls 0º - 180º). In this work, we have run a mesoscale model on the region of Tharsis during the dusty season. We find that due to the different vertical distribution of water in the atmosphere and the higher altitude of the hygropause, the pumping of water by thermal slope winds found by Michaels et al. (2006) is not efficient during this season. Instead, vertically propagating gravity waves perturbing the hygropause is the predominant cloud formation mechanism. We show this result for the AMEC in the early morning, but it also applies to clouds on Tharsis at other local times (not shown).

An Extremely Elongated Cloud over Arsia Mons Volcano on Mars: II. Mesoscale modeling
Hernández-Bernal et al., 2022. Accepted for publication in Journal of Geophysical Research https://doi.org/10.1029/2022JE007352
This is a free version of the manuscript.Future observations that permit higher resolution cloud tracking of the tail are needed to confirm the very fast advection speeds. In addition, spectral observations of the head and the region where the tail extends, will allow the retrieval of parameters of interest for microphysics, such as temperature, water vapor and ice concentration, particle sizes, and condensation nuclei availability (dust). Future modeling efforts should focus on understanding the microphysics enabling the expansion of the AMEC, with the inputs of new observations. Jiang and Doyle (2006) examined two cases of "topographically generated cloud plumes" observed on Earth that visually resemble the AMEC, these plumes are a good terrestrial analogue to the AMEC.

The AMEC is a natural laboratory to learn about the microphysics of the martian atmosphere. The knowledge acquired in this work will be a valuable aid for the planning of new observations that will help improve our knowledge of the AMEC and other orographic clouds on Mars.

**Data Availability Statement**
The GCM and mesoscale model and the configuration details are described in section 2 and references therein. A netcdf file containing the data from the mesoscale model used in figures in this paper is available in: https://doi.org/10.14768/75b44a04-fb39-4a5c-b097-00229d34e5c1

**Acknowledgments**
This work has been supported by the Spanish project PID2019-109467GB-I00 (MINECO/FEDER, UE) and Grupos Gobierno Vasco IT-1366-19. JHB was supported by ESA Contract No. 4000118461/16/ES/JD, Scientific Support for Mars Express Visual Monitoring Camera. The Aula EspaZio Gela is supported by a grant from the Diputación Foral de Bizkaia (BFA). The Applied Physics Department at UPV/EHU supported the stay of JHB in Paris. JHB acknowledges the welcoming hospitality by LMD colleagues during his stay.

The estimated energy consumption associated with this work is 231kWh, including a trip by train from Bilbao to Paris (110 kWh), simulations (110 kWh), and regular computing resources for data analysis and day a day working (11 kWh). 48 kWh were consumed in Spain and 183kWh were consumed in France. The total amount of energy used in this work corresponds to 17.6kg of $CO_2$ and 5200mg of long lived radioactive waste.

# Supporting Information

**Contents of this annex**

Figure S1 to S2

**Additional Supporting Information (Files available in the journal)**

Captions for Movies S1 to S4

**Introduction**

This annex contains Figures S1 to S2, and the caption for the supporting videos.

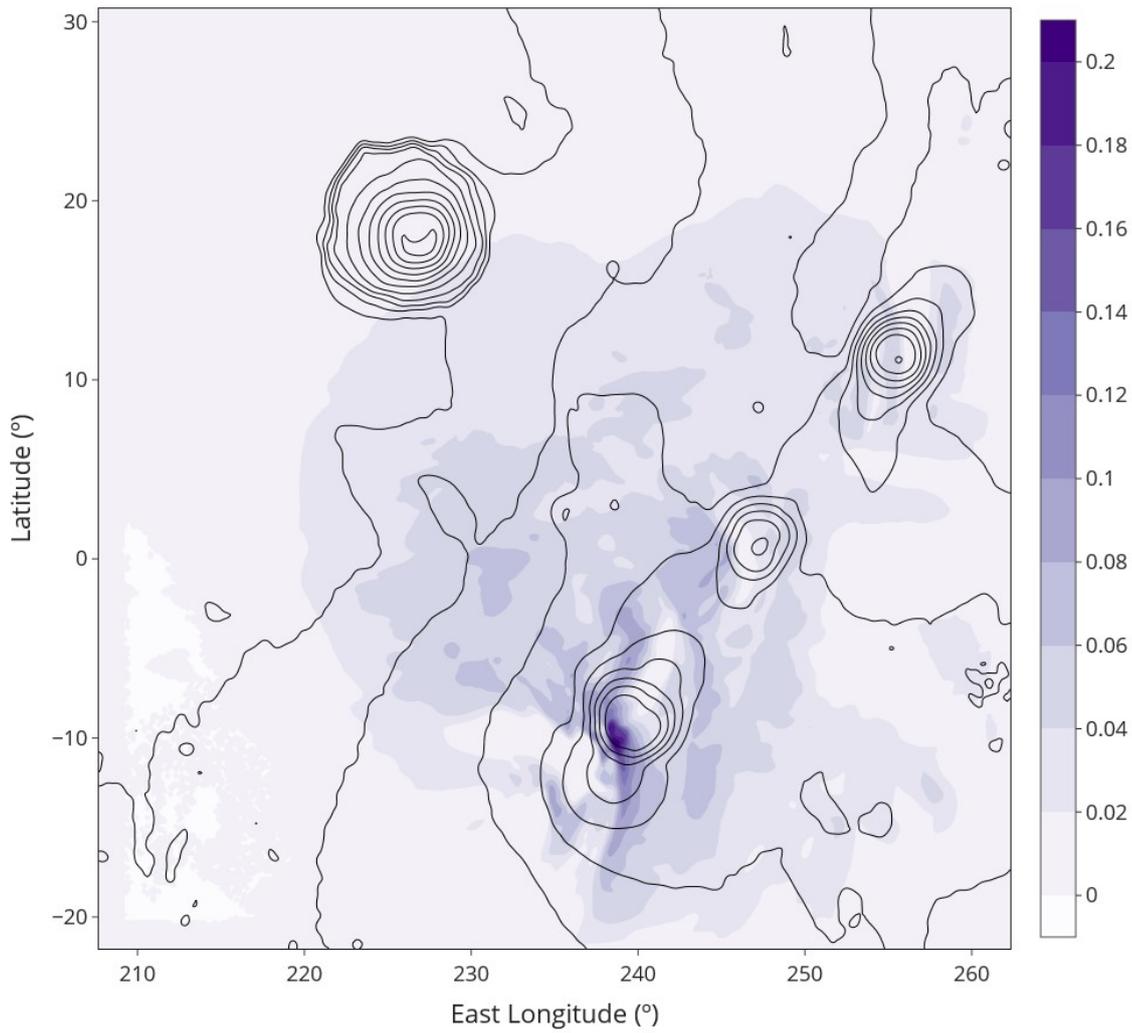

**Figure S1.** This figure is equivalent to Figure 2. Figure 2 is a representation of clouds made from the optical opacity as calculated from the MM, and Figure S1 displays directly the calculated physical optical opacity.

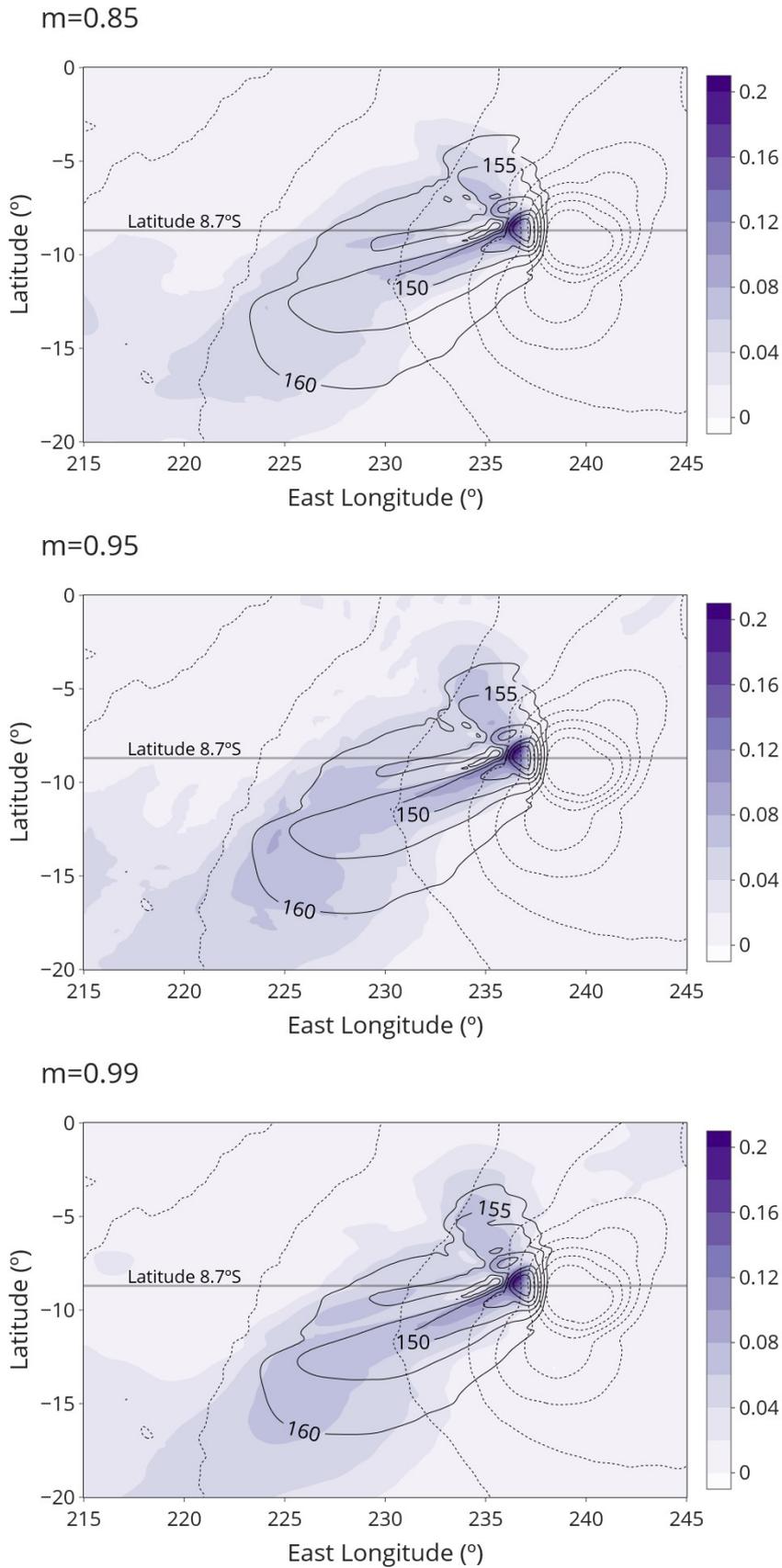

**Figure S2.** Analogue to Fig. 6c for different runs with different contact parameter (m), see section 4.3.3

**Videos S1 and S2**. These videos show cloud formation at two different seasons. S1 corresponds to Ascraeus Mons in Ls 120º, and S2 corresponds to Arsia Mons in Ls 270º. The background shade indicates the total amount of water (including vapor and ice). Thin black contours indicate the amount of water ice, corresponding to condensate clouds. Yellow contours represent areas where the temperature is at least 5K below the mean temperature of the altitude level. And green balls are passive tracers that move following winds (zonal, meridional, and vertical winds). Both videos show slope winds around the volcanoes pumping water (and green tracer balls) to higher levels. Video S1 shows the cloud formation mechanism described by Michaels et al. (2006) for the first half of the year, when the hygropause (the level where water typically condenses) is at low altitude, and any lifting of water is enough to form clouds. Video S2 shows that during the second half of the year, the hygropause is at higher altitudes, and as a result pumped water (traced by green balls) does not reach the hygropause and thus it is not a significant driver of cloud formation. Instead, clouds are correlated to cold pockets around 40 – 50 km in altitude.

**Videos S3 and S4**. These videos are animated versions of Figures 4 and 5 in the main text and we refer to the caption of such figures for description.